\documentclass[conference]{IEEEtran}
\usepackage{amssymb}
\usepackage{amsmath}
\usepackage{bm}
\usepackage{svg}
\usepackage{amsmath}
\usepackage{listings}
\usepackage[hyphens]{url}
\usepackage{multirow}
\usepackage[font={small,it}]{caption}
\usepackage[bookmarks=false]{hyperref} 
\usepackage{algorithm2e}
\usepackage{booktabs}
\usepackage{subfig}
\usepackage{paralist}
\usepackage{float}

\usepackage[left=0.62in,right=0.62in,top=0.75in,bottom=1.0in]{geometry}

\hypersetup{breaklinks=true}
\newcommand{\specialcell}[2][c]{
	\begin{tabular}[#1]{@{}l@{}}#2\end{tabular}}

\makeatother

\newcommand{\name}{\textsc{CoinWatch}\xspace}

\renewcommand{\tablename}{Table}

\renewcommand{\figurename}{Figure}

\newcommand{\myparagraph}[1]{\vspace{0.1cm}\noindent{\it #1.}}

\newcommand{\nameS}{$\mathcal{CW}$\xspace}

\IEEEoverridecommandlockouts

\begin{document}
\title{CoinWatch: A Clone-Based Approach For Detecting Vulnerabilities in Cryptocurrencies
\thanks{$\copyright$ 2020 IEEE.  Personal use of this material is permitted.  Permission from IEEE must be obtained for all other uses, in any current or future media, including reprinting/republishing this material for advertising or promotional purposes, creating new collective works, for resale or redistribution to servers or lists, or reuse of any copyrighted component of this work in other works.}}

\author{
	\IEEEauthorblockN{
		Qingze Hum,\IEEEauthorrefmark{2} Wei Jin Tan, Shi Ying Tey, Latasha Lenus \\
		Singapore University of Technology and Design,\\
		\IEEEauthorrefmark{2}qzhum1996@gmail.com
	}\and
	\IEEEauthorblockN{\hspace{2cm}}\and			
	
	\IEEEauthorblockN{
		Ivan Homoliak~~~~~~~\\
		FIT, Brno University of Technology,~~~~~~~\\
		ihomoliak@fit.vutbr.cz~~~~~~~
	}
	\and
	\IEEEauthorblockN{\hspace{2.cm}}\and			
	\IEEEauthorblockN{
		Yun Lin\\
		National University of Singapore,\\
		dcsliny@nus.edu.sg
	}\and
	\IEEEauthorblockN{\hspace{2cm}}\and			
	\IEEEauthorblockN{
		Jun Sun\\
		Singapore Management University,\\
		sunjunhqq@gmail.com
	}			
}

\maketitle
\begin{abstract}
  Cryptocurrencies have become very popular in recent years.
  Thousands of new cryptocurrencies have emerged, proposing new and novel techniques that improve on Bitcoin's core innovation of the blockchain data structure and consensus mechanism.
  However, cryptocurrencies are a major target for cyber-attacks, as they can be sold on exchanges anonymously and most cryptocurrencies have their codebases publicly available. 
  One particular issue is the prevalence of code clones in cryptocurrencies, which may amplify security threats. 
  If a vulnerability is found in one cryptocurrency, it might be propagated into other cloned cryptocurrencies.
  In this work, we propose a systematic remedy to this problem, called \name (\nameS). 
  Given a reported vulnerability at the input, \nameS uses the code evolution analysis and a clone detection technique for the indication of cryptocurrencies that might be vulnerable. 
  We applied \nameS on 1094 cryptocurrencies using 4 CVEs and obtained 786 true vulnerabilities present in 384 projects, which were confirmed with developers and successfully reported as CVE extensions.
\end{abstract}

\section{Introduction}\label{sec:intro}
 Bitcoin is designed to be a peer-to-peer electronic cash system, introducing an innovative consensus mechanism and incentives for participants. 
 All participants in the consensus protocol must run a full node.
Bitcoin has inspired hundreds of cryptocurrencies utilizing blockchain technology to serve as a decentralized means of transferring crypto-assets.
There were 98 cryptocurrencies introduced in 2016, 412 introduced in 2017, and 726 introduced in 2018. 
Given that Bitcoin's codebase~\cite{bitcoin-codebase} is publicly available, developers may easily fork its repository from GitHub and make custom modifications, which they might later publish as their own cryptocurrency.
In this way, many cryptocurrencies might become popular and adopted by the public.
However, many cryptocurrencies are also ``popular'' due to various high profile security incidents~\cite{SRA-homoliak,bonneau2015sok,wang2018survey,conti2018survey}.

Detecting and patching the vulnerabilities manually is prohibitively costly.
Due to the decentralized nature of cryptocurrencies, whenever a patch is created, individual miners have to patch their own nodes independently.
In the case of the inflation bug in Bitcoin, the vulnerability was fixed by only 31.65\% of all nodes~\cite{inflation-bug-after} even after a full year from the disclosure.\footnote{Note that as of $30^{th}$ September 2020,  the ratio of updated nodes is even worse and is equal to 14.42\%. See \url{https://luke.dashjr.org/programs/bitcoin/files/charts/security.html?201817144}.}
The inflation bug~\cite{inflation-bug} was discovered in September 2018 and allowed the inflation of the total Bitcoin supply.
Upon analysis, it was found that this vulnerability had been present since 2017 but has been patched only in 2019. 
Nevertheless, this vulnerability is shared by 31 cryptocurrencies that emerged as modifications of Bitcoin clones.
Patching the vulnerability across all such cryptocurrencies is of paramount importance for their financial stability.

\begin{figure}[t]
	\centering
	\vspace{0.1cm}
	\includegraphics[width=0.32\textwidth]{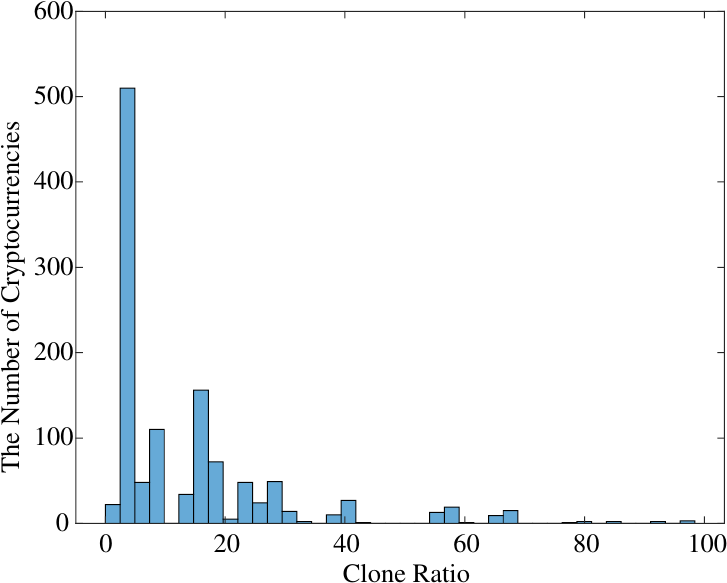}
	\caption{A histogram of forked projects from Bitcoin and their clone ratios as compared to Bitcoin v0.17.0.}
	\label{fig:histogram}
	\vspace{-0.4cm}
\end{figure}
Our empirical study shows that 786 cryptocurrencies were directly or indirectly forked from a version of Bitcoin. 
\autoref{fig:histogram} depicts the histogram of code similarity (i.e., clone ratio) of cryptocurrencies against the Bitcoin codebase.
In most of the cases, the clone ratio is below 30\%; however, some of them have this ratio even higher than $50\%$. 
Therefore, once a vulnerability is discovered in one cryptocurrency, the vulnerability is very likely to be propagated among its clones.
Editing cloned codebases is a common phenomenon in software engineering and maintenance.
Nevertheless, the popularity of forking codebases of cryptocurrencies makes the problem unique in that:
(1) the clones are spread across a vast number of projects, which makes overall clone management difficult,
(2) the code among cryptocurrency projects is of high similarity, and
(3) the financial impact of neglecting a propagated vulnerability might be extremely costly.

In this work, we propose \name (\nameS), a clone-based approach that indicates propagated vulnerabilities across cryptocurrencies.
At the input, \nameS takes  the code fragment relevant to a known vulnerability in a \textit{target} cryptocurrency and the code repositories of \textit{monitored} cryptocurrencies.
First, \nameS analyzes when the vulnerability of the target cryptocurrency was introduced within its incremental code history (e.g., Git).
Next, \nameS goes through the suspicious cryptocurrencies and infers whether they were forked before the vulnerability was introduced.
In the positive case, \nameS conducts clone detection to find out whether the given vulnerability has been already fixed and if not it reports a warning. 

We have built a proof-of-concept tool for \nameS, which to the date scans 1094 cryptocurrencies and compares them to Bitcoin, serving as the target cryptocurrency. 
For evaluation purposes, we selected the top four cryptocurrency-oriented CVEs in terms of severity.
In sum, \nameS reported 786 propagated vulnerabilities across 384 cryptocurrencies, achieving a true positive rate of 89.7\%.
We notified the developers of affected cryptocurrencies and reported vulnerabilities found to NVD NIST. 
In turn, we obtained 4 CVE extensions: 
\textbf{CVE-2018-17144}, \textbf{CVE-2016-10724}, \textbf{CVE-2016-10725}, and \textbf{CVE-2019-7167}, which are described at \url{https://github.com/JinBean/CVE-Extension/}.

\section{An Illustrating Example} \label{sec:example}
In this section, we illustrate how \nameS works while taking CVE-2018-17144 as an example.

\lstset{
    basicstyle=\tiny\ttfamily\scriptsize,
    breaklines=true,
    frameround=tttt,
    frame=trBL,
    basewidth=0.5em,
    linewidth=0.98\columnwidth,
    tabsize=2,
    showstringspaces=false,
    escapeinside={<@}{@>},
}

\subsection{CVE-2018-17144 Explained}
\begin{figure}[b]
	\vspace{-13pt}
	\centering
	\includegraphics[width=0.46\textwidth]{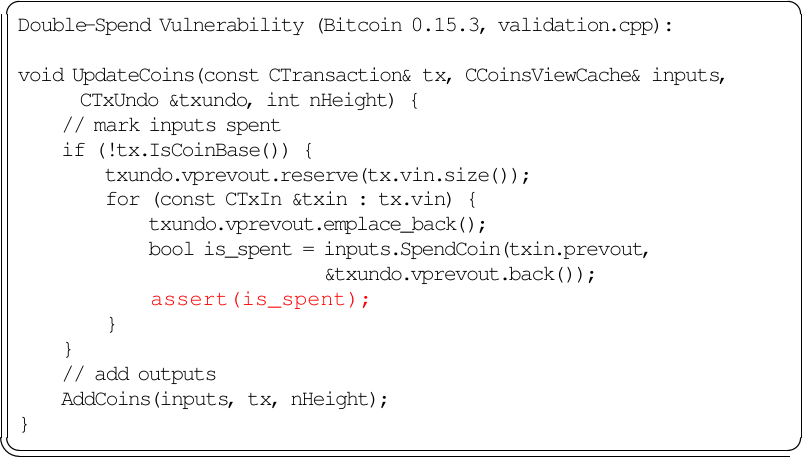}
	\caption{Propagated code of DoS vulnerability (CVE-2018-17144).}
	\label{tbl:double_propagation_a}
		\vspace{-0.1cm}
\end{figure}

This vulnerability is a result of \textit{reachable assertion} weakness CWE-617~\cite{CWE-617}, and it enables the crashing of any node in the network by crafting a transaction that double spends (see \autoref{tbl:double_propagation_a}), which causes a DoS of all nodes that receive a mined block containing such a transaction.
The vulnerability was fixed in Bitcoin release 0.16.3 on Sep 18, 2018. 
We depict the vulnerable period of this vulnerability across various Bitcoin clones in \autoref{tbl:affected_coins_time}.

\subsection{Using \name}
Given the CVE description and details about a vulnerability, the codebase of the target cryptocurrency (i.e., Bitcoin) is analyzed for the specific code segments causing the vulnerability:
\begin{compactenum}
    \item the bug-fixing commit (b8f8019) is identified in GitHub,
    \item the bug-introducing commits (eecffe5, 5083079, 3533fb4) are identified in GitHub,
    \item the first cut of potentially vulnerable cryptocurrencies is generated based on the fork dates of input candidates.
\end{compactenum}
Hence, if a cryptocurrency was created before the bug-introducing commits in its ``parent,'' it is unlikely that it will contain the bug.

In the next step of \nameS, the code segment that patches the vulnerability is identified. 
Given this information, \nameS performs a test using clone detection to check for the existence of the vulnerable code (and the fix) within all monitored cryptocurrencies.
The results mark vulnerable cryptocurrencies: they contain the code with the vulnerability but not the fix.

\begin{table}[t]
	\vspace{0.3cm}
	\footnotesize
	\begin{tabular}{r c c}
		\toprule
		\textbf{Cryptocurrency} & \textbf{Timeframe (MM/DD)} & \textbf{Months Vulnerable} \\ \midrule
		
		Argentum       & 12/16-11/18       & 23                        \\ 
		Chaincoin      & 12/16-10/18       & 22                        \\ 
		CreativeCoin   & 12/16-11/18       & 23                        \\ 
		DigiByte       & 12/16-9/18        & 21                        \\ 
		EliCoin        & 2/18-11/18        & 9                         \\ 
		Irlecoin       & 6/18-present      & not fixed                 \\ 
		Litecoin       & 12/16-9/18        & 21                        \\ 
		Lynx           & 12/6-10/18        & 22                        \\ 
		Machinecoin    & 4/17-10/18        & 18                        \\ 
		MatrixCoin     & 12/16-10/18       & 22                        \\ 
		Methuselah     & 4/18-10/18        & 6                         \\ 
		MinexCoin      & 7/17-4/19         & 21                        \\ 
		MktCoin        & 10/17-present     & not fixed                 \\ 
		PlatinCoin     & 7/18-2/20      & 19                 \\ 
		Quasarcoin     & 11/17-present     & not fixed                 \\ 
		United Bitcoin & 12/16-9/18        & 21                        \\ 
		Unitus         & 4/17-11/18        & 19                        \\ 
		\bottomrule
	\end{tabular}
	\caption{Vulnerable period of CVE-2018-17144 across Bitcoin clones.}
	\label{tbl:affected_coins_time}
	\vspace{-0.4cm}
\end{table}

\subsection{Applying \nameS on CVE-2018-17144}
The full detection process consists of scanning the codebase of all monitored cryptocurrencies (1094) from our dataset while searching for the code that causes the vulnerability. 
On completion, \nameS reports a candidate set of \textit{suspicious} cryptocurrencies that might be vulnerable with a high likelihood.
 
\myparagraph{\textbf{PigeonCoin}}
On 26 September 2018, PigeonCoin was attacked~\cite{pigeoncoin-hack} due to CVE-2018-17144, resulting in a loss of 235 million PGN (i.e., \$15,000 USD).
The fix~\cite{pigeoncoin-fix} of the vulnerability was only uploaded to their GitHub repository on 27 September 2018: 9 days after it was added to Bitcoin. 
If \nameS had monitored the PigeonCoin, the vulnerability would be discovered and fixed before its exploitation.

\section{\name}\label{sec:approach}

\begin{figure*}[t]
	\centering
	\fontsize{7}{10}\selectfont
	\includegraphics[scale=0.75]{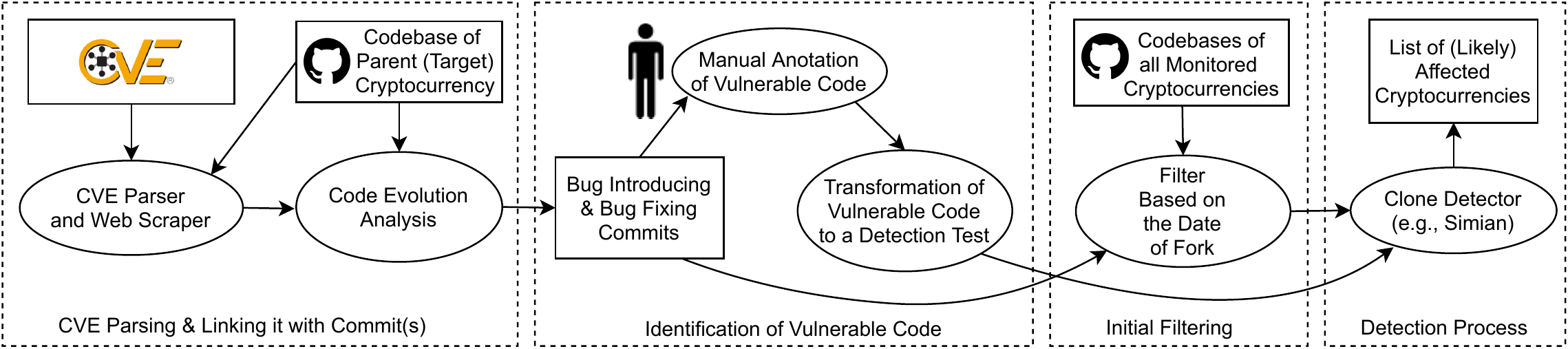}
	\caption{Overview of \name.}\label{tbl:workflow}
	\vspace{-0.4cm}
\end{figure*}

\subsection{Overview}
The overall workflow of \nameS is shown in \autoref{tbl:workflow}. 
First, a target  CVE~\cite{cve} is provided at input together with data publicly obtainable from its structured details.
After picking a target CVE, \nameS performs a code evolution analysis of the parent project to obtain bug fixing and bug introducing commits.
The bug introducing and fixing commits are then manually annotated to minimize the code responsible for the vulnerability and the fix, respectively. 
Then, \nameS generates a detection test for a clone detector.
Before running the clone detector, the list of monitored projects is filtered based on the date of the fork to narrow down the search space by pruning unrelated and unaffected projects.
Then, at the heart of \nameS is the clone detector, which, given the filtered codebases of monitored cryptocurrencies, identifies and reports cloned projects that are likely to be affected by the vulnerability.

\begin{algorithm}[b]
	\small
	\caption{An extended SZZ algorithm.}\label{fig:SZZ}
	
	\SetAlgoLined
	\LinesNumbered
	\SetKwInOut{Input}{Input}
	\SetKwInOut{Output}{Output}
	\Input{~$I$: Issues that match criteria}
	\Input{~$S$: Regex string}
	\Output{~$G_b := \emptyset$: Git commits with bugs}
	\Output{~$G_f := \emptyset$: Git commits with fixes}
	\ForEach {i in $I$}{
		\ForEach {$c$ in i.commits}{
			$G_f$ := $G_f ~\cup~$ RE.match($S$, $c$)}{
			
		}
	}
	\ForEach{cf in $G_f$}{
		$G_b$ := $G_b ~\cup~$ getPrevCommits(cf)\;
	}
	\Return $G_f, G_b$
	\medskip
\end{algorithm}

\subsection{Details}
\myparagraph{\textbf{CVE Parsing}}
After a new vulnerability in the parent project is reported, the \nameS triggers its execution. 
First, \nameS parses the input CVE and extracts details about CVE: 
(1) the date of publishing,
(2) keywords from the description using a text mining for short texts~\cite{carretero2013improving},
(3) external links pointing to resources of the version control system (e.g., issues, pull requests, release notes),
and (4) the list of cryptocurrencies affected by the CVE along with the programming language they were written in.
Then, \nameS performs a check whether the vulnerability is dependent on the code specific to the cryptocurrency since some vulnerabilities may originate from used underlying protocols (e.g., TLS).
Finally, the software release versions corresponding to the bug and the fixes are extracted using web scrapping techniques and utilized for a code evolution analysis that takes into account the incremental code history of a project.

\myparagraph{\textbf{Code Evolution Analysis}} 
\nameS uses the code evolution of a target project (e.g., Bitcoin) to discover the commit IDs containing both the fix and the bug.
\nameS sets the entry point as the release date of the vulnerability and proceeds back in time while iterating over the issues, until it links the fixing commit(s) with the input CVE. 
To track down the commit ID, \nameS utilizes the SZZ algorithm proposed by \textbf{S}liwerski, \textbf{Z}immermann, and \textbf{Z}eller~\cite{sliwerski2005changes}.
SZZ algorithm was developed as an approach to identify bug-introducing commits in a software repository.
By leveraging on the issue-tracking system of GitHub, we improve the accuracy using the \textit{git-blame} command for finding the bug-introducing commits.
An open implementation of the SZZ algorithm called \textit{SZZ Unleashed}~\cite{borg2019szz} provides insight on how to apply it in practice.

\begin{algorithm}[b]
	\small
	\SetAlgoLined
	\LinesNumbered
	\SetKwInOut{Input}{Input}
	\SetKwInOut{Output}{Output}
	\Input{$V_c$: set of monitored projects}
	\Input{$G_b, G_f$: bug-introducing and fixing commits}
	\Output{$V_{vuln}$: set of potentially vulnerable projects}
	\ForEach {$p$ in $V_c$}{
		\If{$p.repoForkDate \geq minDate(G_b)$\ $\wedge $\ $p.repoForkDate \leq maxDate(G_f)$}{
			$V_{vuln} := V_{vuln} ~\cup~ p$\;
		}
	}
	\Return $V_{vuln}$
	\medskip
	\caption{Initial filtering of monitored projects.}\label{fig:pseudocode}
\end{algorithm}
In this work, we built on the SZZ algorithm and extend it for our purpose to identify bug-fixing and bug-introducing commits.
Using regular expression matching (lines 1 to 5 of \autoref{fig:SZZ}, we  search through all issues that have been fixed, resolved, closed, or labeled as a "bug".
\nameS uses keywords such as "CVE", "CVE-ID" and extracted keywords from the CVE details as the string pattern for regular expression matching.
This allows \nameS to locate the bug-fixing commits, which are stored at $G_f$.
To find the bug-introducing commit $G_b$, \nameS uses git-blame functionality, which annotates the line of code with information from the revision that modified the line as the last one (i.e., \textit{getPrevCommits()}).
We emphasize that \nameS takes a conservative approach to pick the oldest date for the bug-introducing commit and the newest date for the bug-fixing commit.
This method allows us to track down both bug-introducing and bug-fixing commits, with which we can estimate the time-frame of the presence of the bug in any codebase.
We use this time-frame as a window representing its propagation within a project that was created by forking the target project.

\myparagraph{\textbf{Identification of Vulnerable Code}}
After obtaining bug fixing and bug introducing commits, \nameS requires manual annotation of the code in these commits (i.e., the code responsible for the vulnerability and the fix).
This requires understanding the cause of the vulnerability; however, this manual effort is only a one-time action (per vulnerability), while all monitored projects are checked automatically later. 
Next, \nameS transforms the annotated code to a detection test suitable for a clone detector.

\myparagraph{\textbf{Filtering of Monitored Projects}}
We provide a simple method for the initial filtering of monitored cryptocurrencies in \autoref{fig:pseudocode}, which is based on the timestamp of forking these projects from the parent project.
Letting $V_c$ be the set of monitored cryptocurrency projects, if $p \in V_c$ is created after $G_b.date$ and before $G_f.date$, we consider $V_c$ for further analysis.
Otherwise, if the repository of $p$ was created before the bug was introduced or after the bug was fixed, we ignore this case since the bug had not existed at the time of forking the project.
The list of candidate projects forms our first cut of potentially vulnerable projects.
\autoref{fig:pseudocode} uses the output of \autoref{fig:SZZ} to identify the dates when the vulnerability was created and presented until by checking the timestamps associated with the particular commit IDs.

\myparagraph{\textbf{Detection Process}}
Finally, using the generated detection tests, \nameS applies the clone detector to find clones of the parent project in the monitored list of cryptocurrencies, which are then analyzed using $G_f$ and $G_b$ to indicate such cloned projects that are potentially vulnerable.

\section{Implementation} \label{sec:implementation}

\myparagraph{\textbf{Experiment Setup}}
All the experiments we performed ran on the machine with a quad-core i7-7500U CPU clocked at 2.70GHz and equipped with 16 GB of memory. 
We conducted a simulation experiment where we applied \nameS on the versions of projects dating before the version where the vulnerability was disclosed. 
The simulation experiment evaluated whether some cryptocurrencies might have been fixed earlier if our approach has been adopted.
To instantiate a clone detector, we chose Simian~\cite{Simian}, a similarity analyzer that identifies duplication in C++ code as well as other tools that use text-based string comparison to identify type I clones.
The Type I clones represent the situation when code fragments in a particular code-base are the same, omitting the comments and indentations~\cite{roy2014vision},~\cite{roy2009detection}.
Note that Type II clones are syntactically similar code fragments, which are usually created from Type I clones by renaming variable names or modifying data types. 
Finally, Type III clones are created due to additions, deletions, or modifications of code fragments in Type I or Type II clones.

\lstset{
	language=XML,
	keywordstyle=\color{red},
	morekeywords={encoding, lineCount, sourceFile, processingTime},
	linewidth=56em,
	xleftmargin=.03\textwidth,
}
\begin{figure*}[h]
	\vspace{0.5cm}
	\centering
	\fontsize{9}{1}
	\selectfont
\begin{lstlisting}
<simian version="2.5.10">
    <check ignoreCharacterCase="true" ignoreCurlyBraces="true" ignoreIdentifierCase="true" ignoreModifiers="true" ignoreStringCase="true" threshold="6">
        <set lineCount="6" fingerprint="5eb7165c145ded7b7469ac9ac53e677b">
            <block sourceFile="\simian\bin\coin\bitcoin-0.16.3\src\test\skiplist_tests.cpp" startLineNumber="24" endLineNumber="31"/>
            <block sourceFile="\simian\bin\coin\Dogecoin\src\test\skiplist_tests.cpp" startLineNumber="25" endLineNumber="32"/>
        </set>
        <set lineCount="6" fingerprint="4ceb8650e14442de03bb7eb1b8f9b9b4">
            <block sourceFile="\simian\bin\coin\bitcoin-0.16.3\src\key.cpp" startLineNumber="203" endLineNumber="211"/>
            <block sourceFile="\simian\bin\coin\Dogecoin\src\key.cpp" startLineNumber="72" endLineNumber="80"/>
        </set>
        <set lineCount="6" fingerprint="7b80d7542d174e7bdb2e532d6a96e1bb">
            <block sourceFile="\simian\bin\coin\bitcoin-0.16.3\src\test\script_tests.cpp" startLineNumber="1282" endLineNumber="1289"/>
            <block sourceFile="\simian\bin\coin\Dogecoin\src\test\script_tests.cpp" startLineNumber="964" endLineNumber="971"/>
        </set>
        ...
    <summary duplicateFileCount="348" duplicateLineCount="34092" duplicateBlockCount="2369" totalFileCount="455" totalRawLineCount="160822" totalSignificantLineCount="94838" processingTime="1624"/>
    </check>
</simian>
\end{lstlisting}
	\caption{An example of XML output from Simian.}
	\label{tbl:simian_output}
		\vspace{-0.3cm}
\end{figure*}

\myparagraph{\textbf{Testing Simian}}
Before we conducted any clone detection experiment, we first ran a test to verify that Simian is capable of accurate detection of cloned  code  within the  full code structure.
Therefore, we performed a positive test and compared a cloned repository of Bitcoin-0.17 with a portion of its own code.  This  identified  the  correct  file  and  line  numbers  of  the entire  code  snippet.
We  then  performed  a  negative  test  with a code that did not exist within Bitcoin-0.17.
We found that to minimize false positives, we had to set the code line threshold to the exact length of each snippet of vulnerable code. 
This ensures that only cryptocurrencies containing the full code causing the vulnerability are detected.
However, doing so increased the chance of false negatives; due to the time-sensitive nature of cryptocurrencies and a large attack surface, we adopted this setting since it is more important to detect truly vulnerable projects than to miss potentially vulnerable ones.

\myparagraph{\textbf{Identifying Vulnerable Code}}
Using the CVE database specific to Bitcoin~\cite{BitcoinVulnerabilities} (i.e., the parent project), we reviewed Bitcoin's CVEs to select vulnerabilities that are contained in its codebase versions in the range $\langle 2016, 2018 \rangle$, a time frame where many alternative cryptocurrencies were created. 
Once a CVE was selected for analysis, \nameS searched for the bug fixing and bug introducing commits. 
Then, we manually annotated the code segments of these commits, which are related to the vulnerability and the fix, based on which, \nameS generated detection tests for Simian.
Next, \nameS parsed information from the Bitcoin version where the vulnerability was introduced in and the version it was eventually fixed at.

\myparagraph{\textbf{Clone Detection in Monitored Projects}}
We wrote Python scripts that scraped \textit{coingecko.com} to collect a dataset of the names of all existing cryptocurrencies, the link of their GitHub repositories, and the language that the cryptocurrency was written in. 
Out of the 1970 cryptocurrencies that we had, 43 were listed with a broken GitHub link or were just links to websites. 
We then used Git to clone all valid repositories. 
Since we wanted to identify vulnerabilities from Bitcoin clones, we removed all cryptocurrencies that had  not used C++. 
We were then left with 1094 cryptocurrencies. 
The folder structure for the analysis is comprised of one main folder, which we will call the \textit{comparison folder}. 
Inside the \textit{comparison folder} were two more folders; the first containing the vulnerable code and the second containing the code of the cryptocurrency we wanted to compare against. 
Python scripts were used to automate the moving of the code from our database into the comparison folder, running the Simian clone detection, and then moving the code back. 
This was done iteratively for each cryptocurrency in the database. 
Simian produced an XML file after every iteration containing the results of the clone detection. 
The XML file contains the file path of each detected clone, the number of lines detected, and the processing time, as seen in \autoref{tbl:simian_output}.
Another script then ran to parse each of the Simian outputs, which gave us the list of suspicious vulnerable cryptocurrencies. 
To support the validity of our methodology, we also experimented with Zcash (and its CVEs) serving as a parent project (see vulnerability CVE-2019-7167).

\myparagraph{\textbf{Responsible Disclosure}}
Finally, before disclosing vulnerabilities in the cloned projects, we filed 4 CVE extensions with MITRE.
Once the CVEs were accepted, we informed the developers of these projects about the vulnerability and the fix recommended in the CVE description.

\section{Evaluation}  \label{sec:evaluation}
We evaluated \nameS through various experiments. 
The experiments were designed to answer the following research questions (RQs):

\begin{compactitem}
	\item \textbf{RQ1}: Are clones prevalent in cryptocurrencies?
	\item \textbf{RQ2}: Can we accurately detect propagated vulnerabilities through clones?
	\item \textbf{RQ3}: Are the propagated vulnerabilities real problems? 
\end{compactitem}

\smallskip\noindent
RQ1 examines our hypothesis, i.e., clones are prevalent in cryptocurrencies, and therefore identifying vulnerabilities by tracking clones is an important problem. 
RQ2 evaluates the accuracy of our approach. 
RQ3 is designed to question the usefulness of \nameS in practice and reports on true positive and false positive rates.

\subsection{RQ1: Clone Prevalence}
To answer RQ1, we systematically applied \nameS to the latest versions of the cryptocurrencies within our dataset scraped from Coingecko as of October 2018.\footnote{https://www.coingecko.com/en} 
In total, there were initially 2079 cryptocurrencies as of Oct 2018, and 4389 cryptocurrencies as of Apr 2019. 
We note that a large number  of new cryptocurrencies were created in this short period of time. 
There are also different types of cryptocurrencies; e.g, 2431 of them are ERC-20 tokens~\cite{EthereumEIP20}, which we consider as out-of-scope.\footnote{Note that ERC-20 tokens are built on the Ethereum platform and contain a standardized smart contract code that is parametrized by particular tokens.}
Out of the remaining 2079 cryptocurrencies that we scraped, 1094 cryptocurrencies are written in C++, the same programming language as Bitcoin is written in. 
The breakdown of the remaining cryptocurrencies can be found in \autoref{tbl:coin_language}.

\begin{table}[b]
	\footnotesize
	
	\centering
	\begin{tabular}{r c} \toprule
		\textbf{Language} & \textbf{The Number of Projects} \\
		\midrule
		C++ &	1094 \\
		Javascript & 334 \\
		C & 65 \\
		Go & 65 \\
		Python & 36 \\
		Java & 30 \\
		Others & 455 \\
		\bottomrule
	\end{tabular}
	\caption{Distribution of programming languages across cryptocurrencies considered in our work.}
	\label{tbl:coin_language}
\end{table}

\setlength{\tabcolsep}{4.2pt}
\begin{table*}[t]
	\vspace{0.3cm}
	\centering
	\footnotesize
	\begin{tabular}{r r c c c c c c c} \toprule
		\textbf{Cryptocurrency} & \textbf{LOC} & \textbf{Bitcoin 0.17.0}           & \textbf{Bitcoin 0.16.3}           & \textbf{Bitcoin 0.15.2}           & \textbf{Bitcoin 0.14.2}           & \textbf{Bitcoin 0.13.2}          & \textbf{Bitcoin 0.12.0}  & \textbf{Bitcoin 0.11.2}           \\ \midrule
		Bitcoin Cash    & 68,545        & 26.0\%           & 32.2\%           & \textbf{34.2\%}  & 27.3\%           & 20.3\%          & 17.2\%  & 13.1\%           \\
		Bitcoin Diamond & 51,102        & 55.6\%           & 70.2\%           & 77.2\%           & \textbf{98.1\%}  & 73.6\%          & 56.2\%  & 42.3\%           \\
		DogeCoin        & 38,073        & 33.2\%           & 40.8\%           & 43.9\%           & 55.2\%           & 67.1\%          & 77.5\%  & \textbf{91.6\%}  \\
		Monero          & 57,232        & 0.0\%            & 0.0\%            & 0.0\%            & 0.0\%            & 0.0\%           & 0.0\%   & 0.0\%            \\
		Zcash           & 51,509        & 21.1\%           & 24.4\%           & 25.1\%           & 33.8\%           & \textbf{42.2\%} & 46.0\%  & 40.7\%           \\
		Dash            & 65,104        & 40.24\%          & 47.23\%          & 52.52\%          & \textbf{63.93\%} & 46.25\%         & 38.62\% & 28.64\%          \\
		DigiByte        & 63,769        & \textbf{90.90\%} & 69.32\%          & 60.81\%          & 42.02\%          & 30.94\%         & 24.14\% & 18.85\%          \\
		Electroneum     & 37,040        & 0.00\%           & 0.00\%           & 0.00\%           & 0.00\%           & 0.00\%          & 0.00\%  & 0.00\%           \\
		EOS             & 96,066        & 0.00\%           & 0.00\%           & 0.00\%           & 0.00\%           & 0.00\%          & 0.00\%  & 0.00\%           \\
		Litecoin        & 57,014        & 76.73\%          & \textbf{98.17\%} & 86.11\%          & 58.55\%          & 43.75\%         & 33.27\% & 25.83\%          \\
		Nano            & 26,987     & 0.00\%           & 0.00\%           & 0.00\%           & 0.00\%           & 0.00\%          & 0.00\%  & 0.00\%           \\
		Qtum            & 112,867       & 37.52\%          & \textbf{48.18\%} & 42.68\%          & 28.82\%          & 21.48\%         & 16.30\% & 12.68\%          \\
		Ravencoin       & 62,200        & 63.33\%          & 81.09\%          & \textbf{82.41\%} & 55.93\%          & 41.22\%         & 31.45\% & 24.27\%          \\
		Steem           & 108,561       & 0.01\%           & 0.02\%           & 0.02\%           & 0.02\%           & 0.02\%          & 0.02\%  & \textbf{0.07\%} 
		\\ \bottomrule
	\end{tabular}
	\caption{Percentage of cloned code as compared to various Bitcoin versions.}
	\label{tbl:percentage_compare}
\end{table*}

As we conjecture that these cryptocurrencies are clones of various versions of Bitcoin, using \nameS, we systematically apply Simian to compare each of these cryptocurrencies with different versions of Bitcoin. 
Part of the results are summarized in \autoref{tbl:percentage_compare}, where the second column shows the overall number of lines of code (LOC) in the project and the following columns show the clone ratio of each cryptocurrency in respect to different versions of Bitcoin. 
The clone ratio is defined as follows:
\begin{eqnarray}
    ratio_{clone} &=&\frac{K}{T},
\end{eqnarray}
where $K$ is the number of LOC cloned from Bitcoin and $T$ is the total number of LOC.
Note that \autoref{tbl:percentage_compare} shows the clone ratio of a subset of 14 cryptocurrencies that were among the top 50 by the market capitalization as of the time of making the experiment. 
The average clone ratio was found to be 30.7\%. 
However, if we compare the highest clone ratio among all versions of the same set of cryptocurrencies written in C++, the average clone ratio increases to 46.6\%. 
Furthermore, multiple cryptocurrencies have a clone ratio of more than 90\%.
Hence, we conclude that code clones are indeed prevalent in cryptocurrencies. 

\subsection{RQ2: Accuracy of \name}
To answer this question, we systematically apply \nameS to multiple vulnerabilities discovered in recent years and see whether they are propagated to other cryptocurrencies. 
We focus on such vulnerabilities due to the following reasons. 
First, these are high-profile vulnerabilities. 
Second, these vulnerabilities are reported between the years 2016 to 2018, in which we evinced a boom of a vast number of cryptocurrencies. 
A vulnerability propagates itself across clones since it is not yet discovered at the time when the parent project is cloned.
The propagation of a vulnerability is highly dependent on when the source code was cloned and how long the vulnerability has existed before it was fixed. 
If a cryptocurrency forked the code during the time frame when the bug was presented in the source code, there is a high chance of the clone containing vulnerable code. 
As a result, we found out that \nameS is capable of accurate detection of vulnerabilities in cloned projects as proven by filing 4 CVE extensions.

\begin{table}[t]
	\centering
	\begin{tabular}{@{}rcccc@{}}
		\toprule
		\textbf{Vulnerability}                      & \textbf{\specialcell{Parent\\Project}} & \textbf{\specialcell{\# Investigated\\~~~~Projects}} & \multicolumn{1}{c}{\textbf{TPR}} & \multicolumn{1}{c}{\textbf{FPR}} \\ \midrule

		\textbf{CVE-2018-17144}                 & Bitcoin &  31   &  100.0\%                                  &     0.00\%                             \\ 
		
		\textbf{CVE-2016-10724}       & Bitcoin   & 422           &  89.3\%                           &    10.7\%                              \\

		\textbf{CVE-2016-10725}                & Bitcoin & 422     &    89.3\%                              &         10.7\%                         \\		
		
		\textbf{CVE-2019-7167} & Zcash  & 1 & 100.0\%     & 0.00\%  \\
		
		\bottomrule
	\end{tabular}

	\caption{A summary of the results obtained by \nameS.}\label{tab:summary-results}
\end{table}

\myparagraph{\textbf{Case Study}}
In the following, we focus on CVE-2018-17144. 
Among 985 pre-filtered cryptocurrencies monitored using \nameS, we found that as of October 2018, 31 cryptocurrencies had Type I clones of a vulnerable code found in Bitcoin and hence were exposed to this vulnerability. 
We monitored the situation after we alerted vulnerable projects and our CVE extension was published at NVD. 
In February 2019, the results from \nameS suggest that 22 of the 31 cryptocurrencies have patched the vulnerability, while in June 2020 there were still 4 projects unfixed.
This implies that attackers had at least a month's time frame from the time the CVE was announced in September 2018 to attack the 31 vulnerable cryptocurrencies, exposing up to $\$35,585,594$ market capital (as of 07 October 2018).
Moreover, in February 2019 and June 2020, attackers could still attack the remaining 9 and 4 unfixed cryptocurrencies, respectively.

\subsection{RQ3: Is a clone equal to a vulnerability?}\label{sec:RQ3}
To answer this question, we investigate how many of the candidate cryptocurrencies that are discovered using \nameS are indeed vulnerable. 
This is a challenging problem since we must be able to determine whether the affected code is reached or not, and whether its execution would result in the same reported vulnerability as in the target project.  
Therefore, we attempt to answer the question in two ways. 
First, we manually examined all the patches of 4 CVEs (that we extended) and cross-checked them with cryptocurrencies reported  by \nameS. 
For each patch, we checked if the vulnerable cloned code was already modified (by maintainers) in the most recent version of the project, and if so, whether it was modified in a way similar to the fix of the CVE in the parent project. 
In the positive case, we considered a discovery made by \nameS on a particular cryptocurrency as a true vulnerability.
The second way we answer the question was manual checking with the developers/maintainers of the project.

\myparagraph{\textbf{True Positives VS False Positives}}
The summary of our results is presented in \autoref{tab:summary-results}.
In the case of CVE-2018-17144, due to an extremely severe consequence of this double-spending vulnerability, many popular cloned cryptocurrencies quickly applied the fix after the CVE was released, e.g. Dash, Ravencoin, Bitcoin Cash, etc. 
This is not the case for low-profile vulnerabilities or for smaller cryptocurrencies, which are less diligent in keeping their codebases updated. 
Pigeoncoin is an example of a cryptocurrency that suffered from this vulnerability, and which was also exploited. 
We analyzed different releases of Pigeoncoin and found that it contained the double-spending vulnerability from its official launch on 07 April 2018 until 27 September 2018. 
The reason for the bug fix on 27 September was the attack made on it. 
If \nameS were to monitor Pigeoncoin at the time the Bitcoin patch was released, we could have identified the vulnerability and prevented the attack. 
In the case of CVE-2018-17144 and CVE-2019-7167, there were no false positives since all candidate cryptocurrencies reported were confirmed to contain these vulnerabilities.

On the other hand, \nameS reported a few false positives as well.
For example, although we initially found out that 422 cryptocurrencies had enabled the alert system (causing CVE-2016-10724 and CVE-2016-10725 in the parent project), 45 cryptocurrencies reported at the output of \nameS did not reference it during compilation and thus are not vulnerable.
Due to this, the true positive rate of \nameS was deteriorated to 89.3\% for these particular vulnerabilities.

\lstset{
	basicstyle=\tiny\ttfamily\scriptsize,
	breaklines=true,
	frameround=tttt,
	frame=trBL,
	basewidth=0.5em,
	xleftmargin=.02\textwidth, xrightmargin=.615\textwidth,
	tabsize=1,
	showstringspaces=false,
	escapeinside={<@}{@>},
}
\begin{figure}[t]
	\vspace{2pt}	
	\centering
	\includegraphics[width=0.46\textwidth]{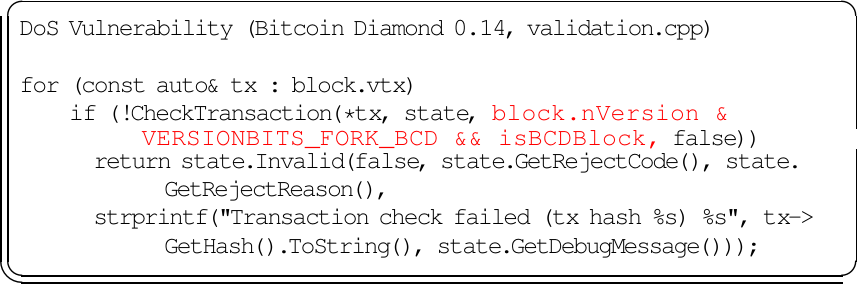}
	\caption{Propagated code for DoS vulnerability in Bitcoin Diamond.}
	\label{tbl:bcd_dos}
\end{figure}
\begin{figure}[t]
	\centering
	\includegraphics[width=0.46\textwidth]{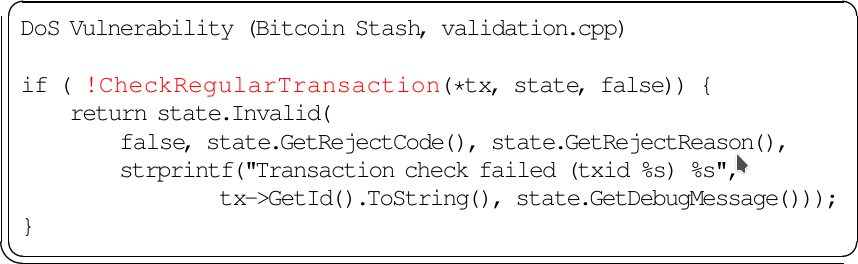}
	\caption{Propagated code for DoS vulnerability in Bitcoin Stash.}
	\label{tbl:bitstash_dos}
	\vspace{-0.6cm}
\end{figure}

\myparagraph{\textbf{Type II and III Clones and Possible False Negatives}}
Type II and III clones come into play when the codebase is edited by changing the way certain functions work or are named. 
One example of a Type III clone that we identified was presented in Bitcoin Diamond Version 0.14.0 and 0.13.0. 
The code can be seen in \autoref{tbl:bcd_dos}. 
Bitcoin Diamond modified its function \textit{CheckTransaction()} to include new arguments that are specific to Bitcoin Diamond itself. 
An example of Type II clone was identified in Bitcoin Stash as seen in \autoref{tbl:bitstash_dos}, where the \textit{CheckTransaction()} function was renamed to \textit{CheckRegularTransaction()}. 
These clones cannot be detected using Simian as the clone detector, and they require more sophisticated tools which we will consider in our future work.
We contacted the development team of these two projects regarding the possible vulnerabilities but did not get the response; hence, we cannot confirm whether these two cases represent false negatives.

\section{Discussion}

\myparagraph{\textbf{Scope of Clones}}
The results presented in this paper are related mainly to Bitcoin, as it is currently the dominant cryptocurrency and has the most number of forks from its GitHub repository.
However, there are other cryptocurrencies written in different programming languages (or are close-sourced) that are similar to Bitcoin but would not be detected as Type I clones.
There is a trade-off between false positives and false negatives if we use more complex types of code clone detection techniques such as the ones searching for Type II or Type III clones.
In some cases we would have to conduct a decompilation of binaries and do the analysis on abstract syntax trees of assembly rather than a verification of source code for match.
This is the direction that we will investigate in our future work.

\myparagraph{\textbf{Manual Annotation}}
A semi-automated nature of \nameS is another limitation, which stems from the manual annotation of the vulnerable code and fix within the bug fixing and bug introducing commits. 
However, this is a one-time action per vulnerability and can be performed in a relatively short time, while later \nameS continues automatically in checking the full list of monitored projects.
In our future work, we plan to design simple heuristic methods that can automate this step.

\myparagraph{\textbf{Non-Standard Assumptions}}
\nameS assumes that if a cryptocurrency project was forked after the bug was introduced in the parent project and before the bug was fixed, it is likely to suffer from the vulnerability. 
However, we note that sometimes forked projects may cherry-pick commits from their parents or apply patches mitigating some security issues, which we currently do not distinguish in \nameS.
However, we emphasize that such selected projects represent only the first cut of potentially vulnerable candidate projects and if they contain a fix, the clone detector will mark them as safe.
In the worst case, if the fix is presented in a modified form that is not matched by the clone detector, they are reported as false positives -- this causes a notification to the maintainers, who just ignore it.

\section{Related Work}\label{sec:related_work}
\textbf{Security of Cryptocurrencies.}
The security of cryptocurrencies has been undermined by several high profile incidents. 
Bitcoin~\cite{nakamoto2008bitcoin} itself has been described to be vulnerable to several attacks such as double-spending~\cite{karame2012double}, transaction malleability ~\cite{decker2014bitcoin}, networking attacks ~\cite{heilman2015eclipse}~\cite{apostolaki2017hijacking}, attacks targeting mining~\cite{eyal2018majority}~\cite{sapirshtein2016optimal}~\cite{eyal2015miner}, and others~\cite{SRA-homoliak}~\cite{bonneau2015sok}.
As a research topic, Bitcoin's security is attractive due to a broad number of use cases proposed for its underlying technology, blockchain.
Bitcoin has inspired several cryptocurrencies to build and improve different aspects of its code to increase its transaction throughput (eg. Bitcoin Cash) or provide greater anonymity (eg. Bitcoin Private).
The intense scrutiny of Bitcoin's security has led to a gap in research on the effects of vulnerabilities from Bitcoin propagating to other cryptocurrencies that have copied parts of Bitcoin's code.
	These cryptocurrencies also store value but are not as valuable as Bitcoin and thus do not have as many developers and resources to build a body of research around their security.
	Therefore, we developed \nameS\ to address the security problem of propagated vulnerabilities from Bitcoin while focusing on real-world attacks that have been identified and fixed within Bitcoin's code.
\\
\textbf{Clones among Cryptocurrencies.}
	Some cryptocurrencies fork existing repositories and launch their own cryptocurrencies with a specific focus.
	As these cryptocurrencies are changing only a portion of the forked codebase before launching, a large portion of their code is cloned from the specific version it was forked from.
	Prior work suggests that code clones increase maintenance effort~\cite{monden2002software}~\cite{mondal2012comparative}, might cause bugs (Li et. al analyzed Linux, FreeBSD, Apache~\cite{li2006cp}), and signal errors in other parts of a codebase.
	It is known that clone genealogy information can be used to identify clones that may benefit new approaches and clone management~\cite{kim2005empirical}.
	Based on the findings of the authors of~\cite{kim2005empirical}, aggressive code refactoring might not be necessary for code clones that are volatile and that the technique of refactoring needs to be complemented by other code maintenance approaches.
	Prior research analyzing package dependencies~\cite{zhang2015assessing} utilizes  a systematic approach of measuring the attack surface exposed by individual vulnerabilities through component level dependency analysis.
	In this work, cloned cryptocurrencies do not utilize Bitcoin as a package but the concept that the attack surface is dependent on code utilized from elsewhere. 
	The authors of~\cite{reibel2019short} analyzed code diversity in the cryptocurrency projects basing on the source code similarity. 
	The intention of the authors was to  examine the extent to which new cryptocurrencies provide innovations.

\textbf{Code Clone Detection.} Basic code clone detection techniques are performed using several tools and generally follow a 6-step process~\cite{roy2009comparison} to generate clone pairs.
Code cloning can be used for the detection of code refactoring, greater efficiency of development, etc.
	However, this methodology has been used for other purposes due to the large volume of open-source code that can be analyzed, such as detecting vulnerabilities.
	For example, Viertel et al.~\cite{viertel2019detecting} utilized this technique for detecting vulnerabilities in codebases of Java-written projects, in a similar fashion than \nameS does. 
	A similar environment is represented by smart contracts in Ethereum or other smart contract platforms.
	Early work focused on finding semantic clones in smart contracts that contained vulnerabilities in the smart contract logic is presented in~\cite{liu2018eclone}.
	A tool detecting previously discovered smart contract vulnerabilities through clones was proposed, and can quickly identify other smart contracts that are semantically equivalent (and thus might contain the same set of vulnerabilities).
	The automatic classification of Ponzi smart contracts based on code clones was proposed in~\cite{chen2018detecting}, and it bases on the fact that related smart contracts have fundamentally similar programming logic.
	Clone detection was also performed after transferring the code into an abstract form such as strings~\cite{johnson1994substring}, tokens~\cite{higo2002software}, and abstract syntax trees (AST)~\cite{koschke2006clone}.
	An efficient heuristic technique basing on edit distance and AST for pair-wise comparison of smart contract byte codes was proposed in~\cite{hartel2019empirical}.
	Techniques such as line comparison based on dot plots~\cite{rieger2005effective}, comparing whole files~\cite{manber1994finding}, pattern recognition of code characteristics through clustering~\cite{merlo2004linear}, finding clones within a syntactic unit through post-processing ~\cite{higo2002software}, and pre-processing~\cite{cordy2004practical} are also used for clone detection.

\section{Conclusion}\label{sec:conclusion}
In this work, we proposed \name, an approach that monitors a  specified set of cryptocurrency projects for the presence of vulnerabilities disclosed in their parent projects.
We applied \name on 1094 monitored cryptocurrencies using 4 CVEs at the input, and we identified 786 true vulnerable projects, which were confirmed with developers and successfully reported as CVE extensions.
Our results show that many attacks could have been prevented should \name has been monitoring the attacked projects. 

In future work, we plan to extend code clone detection to Type II and III clones to increase the detection performance as well as eliminate a need for a manual step of annotating the code within bug fixing and introducing commits.

\bibliographystyle{IEEEtran}

\IEEEtriggeratref{34}
\bibliography{ref.bib}

\end{document}